\documentclass[12pt]{article}
\usepackage{graphicx}
\usepackage{amsmath}
\usepackage{amssymb}

\def\gtap{\raisebox{-.4ex}{\rlap{$\sim$}} \raisebox{.4ex}{$>$}}

\def\sinthl {${\rm sin}^2 \theta_W^{\ell\ {\rm eff}}$}

\def\alr {$A_{LR}$}
\def\alrsp {$A_{LR}\ $}

\def\afbbsp {$A_{FB}^b\ $}

\def\mhsp {$m_H$\ }
\def\chisqsp {$\chi^2$\ }

\def\journal{\topmargin 0.0in   \oddsidemargin 0in
        \headheight 0pt \headsep 0pt
        \textwidth 6.5in 
\textheight 9in 
        \marginparwidth 1.5in
        \parindent 2em
        \parskip .5ex plus .1ex         \jot = 1.5ex}
\journal

\begin{document}
\begin{titlepage}


\begin{center}
\vskip .5in {\large Electroweak Constraints on the Fourth Generation at 
  Two Loop Order}
\vskip .5in
Michael S. Chanowitz
\vskip .2in

{\em Theoretical Physics Group\\
     Lawrence Berkeley National Laboratory\\
     University of California\\
     Berkeley, California 94720}
\end{center}

\vskip .25in

\begin{abstract}

If the Higgs-like particle at 125 GeV is the standard model Higgs
boson, then SM4, the simplest four generation extension of the SM, is
inconsistent with the most recent LHC data. However, 4G variations
(BSM4) are possible if the new particle is not the SM Higgs boson
and/or if other new quanta modify its production and decay rates.
Since LHC searches have pushed 4G quarks to high mass and strong
coupling where perturbation theory eventually fails, we examine the
leading nondecoupling EW (electroweak) corrections at two loop order
to estimate the domain of validity for perturbation theory. We find
that the two loop hypercharge correction, which has not been included
in previous EW fits of 4G models, makes the largest quark sector
contribution to the rho parameter, much larger even than the nominally
leading one loop term. Because it is large and negative, it has a big
effect on the EW fits. It does not invalidate perturbation theory
since it only first appears at two loop order and is large because it
does not vanish for equal quark doublet masses, unlike the one loop
term.  We estimate that perturbation theory is useful for $m_Q\simeq
600$ GeV but begins to become marginal for $m_Q\, \gtap\, 900$
GeV. The results apply directly to BSM4 models that retain the SM
Higgs sector but must be re-evaluated for non-SM Higgs sectors.

\end{abstract}

\end{titlepage}

\newpage

\renewcommand{\thepage}{\arabic{page}} \setcounter{page}{1}

{\noindent \large \bf{1. Introduction}} 
\vskip 0.1 in 

If the Higgs-like particle at 125 GeV is actually the standard model
Higgs boson, then SM4, the simplest four generation extension of the
standard model, is disfavored, first just by its small mass which
exacerbates the stability\cite{hashimoto} and little hierarchy
fine-tuning\cite{holdom_lhp} problems and second by the most recent
production and decay data\cite{lenz_no}. However, the virtues of a
fourth generation remain\cite{4G_virtues} and the problems of SM4
might be addressed in BSM4 models that introduce additional new quanta
and/or non-SM Higgs sectors. For instance, although $m_h\, \gtap\,
m_Q$ is necessary for vaccuum stability in SM4, two Higgs doublet 4G
models can have stable vacua with a light Higgs, $m_h<m_Q$, even for
masses as light as $\sim 100$ GeV\cite{hashimoto}. Two Higgs doublet
4G models can also be consistent with current LHC data for the 125 GeV
state\cite{soni, he-chen, he-valencia} and with EW
data\cite{2dhmG4_EW}. In addition, other new quanta could change loop
mediated Higgs production and decay amplitudes and ameliorate the
stability and little hierarchy problems.

The fourth generation can also play a role in scenarios in which the
Higgs-like particle is not elementary. If it is a strongly bound
composite state\cite{mchm}, fourth generation TeV scale fermions could
be the substrate matter fields on which the new strong dynamics
acts. For instance, models have been constructed in which conformally
invariant strong dynamics acting on the fourth generation breaks
electroweak symmetry\cite{pq123}, engendering two composite Higgs
doublets that mix with an elementary doublet\cite{pq4}. Similarly, if
it is a pseudo-dilaton generated by approximately conformal high
energy dynamics\cite{dilaton_refs}, a TeV scale fourth generation
could again provide the matter field substrate. In this case there
must be a strongly interacting symmetry breaking sector at the TeV
scale, either Higgsless or with a heavy Higgs boson, and the 4G
fermions can provide the necessary oblique corrections to ensure
consistency with the EW data\cite{ew_sm4, hashimoto, ew_ckm4}. The
psuedo-dilaton and heavy Higgs boson could mix, as in a similar 4G
scenario with Higgs-radion mixing\cite{radion}.

There is then no no-go theorem that definitively excludes a
sequential fourth generation once our horizon expands beyond
SM4. The situation is not unlike that of TeV-scale supersymmetry,
which is also disfavored in its minimal (MSSM) version but is not
ruled out in a variety of variations. In both cases, direct searches
for the associated heavy quanta, 4G fermions or SUSY partners, should
be pursued to the limits of the LHC's capability.

LHC searches for the fourth generation $T$ and $B$ quarks have already
pushed the 95\% exclusion limits to $m_B > 611$ GeV\cite{cmsB} and
$m_T > 656$ GeV\cite{atlasT}, assuming $B\to tW$ and $T\to bW$ are the
dominant decay modes. If the 4G quarks are stable or decay outside the
detector the limits are even stronger: using the cross section
corresponding to the 737 GeV lower limit on long-lived stop
production\cite{stop_limit}, we find a 95\% lower limit, $m_T > 930$
GeV.\footnote{We obtained this result from Madgraph\cite{madgraph}
  with a $K$ factor for $gg \to \overline QQ$ of
  1.3\cite{QQ_Kfactor}.} \footnote{Bound state
  formation\cite{bound_states} and/or alternative decay
  scenarios\cite{alt_decays} could allow the 4G quarks to evade the
  above limits.} These limits exceed the so-called ``unitarity limit''
at $m_Q \simeq 500$ GeV for which the leading partial wave amplitude
for $\overline QQ$ scattering saturates unitarity in tree
approximation\cite{cfh}, due to the strong Yukawa couplings of the
quarks. This raises the question that motivated the work presented
here: can perturbation theory be used to obtain the precision
electroweak constraints on the properties of such heavy 4G quarks? In
the process of addressing this question we became aware of the
importance of the two loop hypercharge correction, that is relevant
and important even within the perturbation theory domain of validity.

While sometimes referred to as a unitarity limit, the 500 GeV scale is
not a limit on the allowed masses but is just a landmark for the onset
of strong coupling in $\overline QQ$ scattering. It does not precisely
indicate where perturbation theory fails in other processes, such as
in the corrections to the EW data. As in QCD near nonperturbative
boundaries, reliability can only be estimated process by process, from
the magnitude of the higher order corrections in each case. We do so
here by including the leading nondecoupling two loop corrections. We
find that the expansion is under reasonable control at $m_Q=600$ GeV,
where the two loop corrections to $\rho$ are roughly $\sim 10\%$ as
large as the one loop terms, growing to $\sim 18\%$ at $m_Q=750$ GeV
and $\sim 25\%$ at $m_Q=900$ GeV. For leptons tree unitarity is
saturated at a higher scale, $m_{L,N}\simeq 1$ TeV.\cite{cfh} Since
the electroweak fits typically prefer much lighter lepton masses,
convergence for large lepton masses is a less pressing issue.

However, even when the perturbation expansion is a useful
approximation, for instance, at $m_Q=600$ GeV, the one and two loop
fits yield very different results. This does not signal a breakdown of
perturbation theory because it is due to the hypercharge correction,
which breaks the custodial $SU(2)$ even for equal $T$ and $B$ quark
masses, and only first occurs at two loop order. To assess convergence
we need the three loop hypercharge correction, which has not been
computed. We use a conservative generic estimate of the magnitude of
the three loop hypercharge correction, which is consistent with the
ratio of the one and two loop non-hypercharge contributions, to
estimate the effect of the uncertainty it generates.

Although not surprising with hindsight, it at first seems surprising
that the two loop hypercharge correction is bigger, and even much
bigger in the region of parameter space preferred by the EW fits, than
the nominally leading one loop quark sector
contribution. Nondecoupling contributions to $\rho$, proportional to
the square of the heavy fermion masses, require breaking of the
custodial $SU(2)$ symmetry that preserves $\rho=1$. These arise at one
loop order from mass splitting within the quark or lepton doublets, as
in the well known top quark contribution to the $\rho$
parameter\cite{veltman,cfh}. Since hypercharge breaks custodial
$SU(2)$, it gives rise at two loop order to a nondecoupling correction
to $\rho$, proportional to $g^{\prime\, 2}m_Q^2$, even if the weak
doublet fermions are degenerate. As discussed in section 3, the
interplay between the lepton and quark contributions to the oblique
parameters $S$ and $T\cite{pt}$ favors suppression of the one loop
quark contribution to $T$ and enhancement of the two loop hypercharge
correction in the parameter region preferred by the fits.  Because the
hypercharge correction to $\delta\rho=\alpha T$ is large and negative,
it offsets positive contributions from quark and lepton doublet mass
splitting and from CKM4 mixing\cite{ew_ckm4}. Because of the
correlation with $S$ the preferred region has large $L-N$ and small
$T-B$ mass splitting. We use the analytic expressions obtained by van
der Bij and Hoogeveen\cite{vdbh} for both the
O($(m_T^2-m_B^2)m_{Q}^2)$ and O($g^{\prime\, 2}m_Q^2$) two loop
corrections.

Our fits incorporate a new two loop result\cite{new_rb} for
$R_b=\Gamma(Z \to \overline bb)/\Gamma(Z \to {\rm hadrons})$ that
causes the p-value of the 3G SM fit to fall to 5\% (see section 2).
For this study we assume negligible CKM4 mixing, at or
below the few percent level, for which 4G corrections are fully
captured by $S$ and $T$. Like the oblique fit, the \chisqsp minima for
the best SM4 fits are typically $\sim 2$ units lower than for SM3. The
SM4 and SM3 fits then have comparable p-values, since the SM4 fits,
being oblique, have effectively two additional degrees of freedom.
The two loop SM4 fits have lower \chisqsp minima than the one loop
fits, by $\sim 1/2$ to 3/4 units, since the large negative
contribution to $T$ from the hypercharge correction allows the two
loop fit to approach more closely to the \chisqsp minimum of the
oblique fit, as discussed below.

The more important difference between the one and two loop fits is
in the predictions for the 4G masses. As an example of how the fits
would be used in practice, we imagine a scenario in which the masses
of the $T,B$ quarks and charged lepton $L$ are known and use the EW
fit to constrain the mass of the neutrino $N$. The resulting 
differences between the one and two loop fits are substantial, even
for masses for which the perturbation expansion is under control. We
also show how the constraint on $m_N$ is affected by the uncertainty
in the magnitude of the two loop corrections as a function of the
quark masses.

In section 2 we summarize the current status of the SM3 and oblique
fits, to establish baselines for the SM4 fits. In section 3 we present
the fits with the leading two loop nondecoupling corrections and
compare them to the one loop fits. In section 4 we estimate the effect
on the EW fit of the uncertainty in the perturbation expansion as a
function of the 4G quark masses. Section 5 is a brief discussion of
the results.

\begin{table} [t!]
\begin{center}
\vskip 12pt
\begin{tabular}{c|c|cc|cc}
\hline
\hline
 &Experiment& {\bf SM} & Pull &{\bf Oblique} & Pull\\
\hline
$\Delta \alpha^{(5)}(m_Z)$ & 0.02750 (33) &0.02739& 0.3 & 0.02754 & -0.1\\
$m_t$ & 173.2 (0.9) &173.3  &-0.09&173.3  &-0.1 \\
$\alpha_S(m_Z)$ &    &0.1186& &0.1180 \\
\hline
$S$ & & & & 0.05 & \\
$T$ & & & & 0.08 & \\
\hline
$A_{LR}$ & 0.1513 (21)  & 0.1476  & 1.8& 0.1473 & 1.9 \\
$A_{FB}^l$ & 0.01714 (95) &0.01633  & 0.8&0.01627 &0.9 \\
$A_{e,\tau}$ & 0.1465 (33) & 0.1476 & -0.3& 0.1473 & -0.3 \\
$A_{FB}^b$ & 0.0992 (16) & 0.1034 & -2.7&0.1033&-2.5 \\
$A_{FB}^c$ & 0.0707 (35) & 0.0739 & -0.9&0.0738&-0.9 \\
$Q_{FB}$ & 0.23240 (120) & 0.23145 & 0.8&0.23149&0.8 \\
$\Gamma_Z$ & 2495.2 (23) & 2495.5 &-0.1& 2496.7 & -0.7 \\
$R_{\ell}$ & 20.767 (25) &20.742  & 1.0 &20.737 & 1.2 \\
$\sigma_h$ & 41.540 (37) & 41.478 &1.7 & 41.481 &1.6 \\
$R_b$ & 0.21629 (66) & 0.21475 &2.3& 0.21475 &2.3 \\
$R_c$ & 0.1721 (30) & 0.1722 &-0.05& 0.1722 &-0.04 \\
$A_b$ & 0.923 (20) & 0.935 &-0.6 & 0.935 &-0.6 \\
$A_c$ & 0.670 (27) &  0.668 & 0.08&  0.668& 0.08 \\
$m_W$ & 80.385 (15) & 80.365 & 1.3 & 80.383 & 0.1 \\
\hline
$\chi^2$/dof& & 22.2/13 && 20.3/11 \\
CL($\chi^2$/dof) & & 0.05 &&0.04 \\
\hline
\hline
\end{tabular}
\end{center}
\caption{\small{SM and oblique fits for \mhsp= 125 GeV compared to 
    winter 2012 EWWG data\cite{ewwg}.}}
\end{table}

\vskip 0.1 in
{\noindent \large \bf{2. Standard Model and Oblique Fits}}
\vskip 0.1 in

We use the data set and methods of the Electroweak Working
Group\cite{ewwg}. The SM radiative corrections are computed with
ZFITTER\cite{zfitter}, including the two loop SM electroweak
contributions to $m_W$ and \sinthl. The largest experimental
correlations are included, taken from the EWWG. We use the EWWG data
set with one exception: we do not include $\Gamma_W$, the
$W$ boson width, because it is much less precise than the other
measurements and has no discernable impact on the output parameters.
The resulting SM and oblique fits with \mhsp fixed at 125 GeV are shown
in table 1. 

The new $R_b$ calculation causes the \chisqsp of both fits to increase
by $\sim 5$ units and the p-value of the SM fit to fall to 5\%. In
addition to $R_b$ the major contributor to the \chisqsp is, as in the
past, the conflict between \afbbsp and \alr, which was the principal
cause of the marginal 16\% p-value of the fit using the previous $R_b$
calculation. This 95\% exclusion should be taken seriously, since it
is not diluted by a ``look elsewhere'' effect but is a valid
statistical indicator of the likelihood that the outliers in the fit
could have arisen by statistical fluctuations. We should then look
either to systematic error or new physics as the most likely
explanation.  Systematic error could be theoretical or experimental,
and a leading possibility is the use of a hadronic Monte Carlo to
assess the effect of gluon radiation on the \afbbsp
measurement\cite{msc_ggi}. With slight modifications most new physics
scenarios that addressed the \afbbsp -- \alrsp conflict can also
incorporate the new $R_b$ result. See \cite{ltw-rb} for a model
addressing the current fit with references to the earlier literature.

\begin{figure}
\centerline{\includegraphics[width=3in,angle=90]{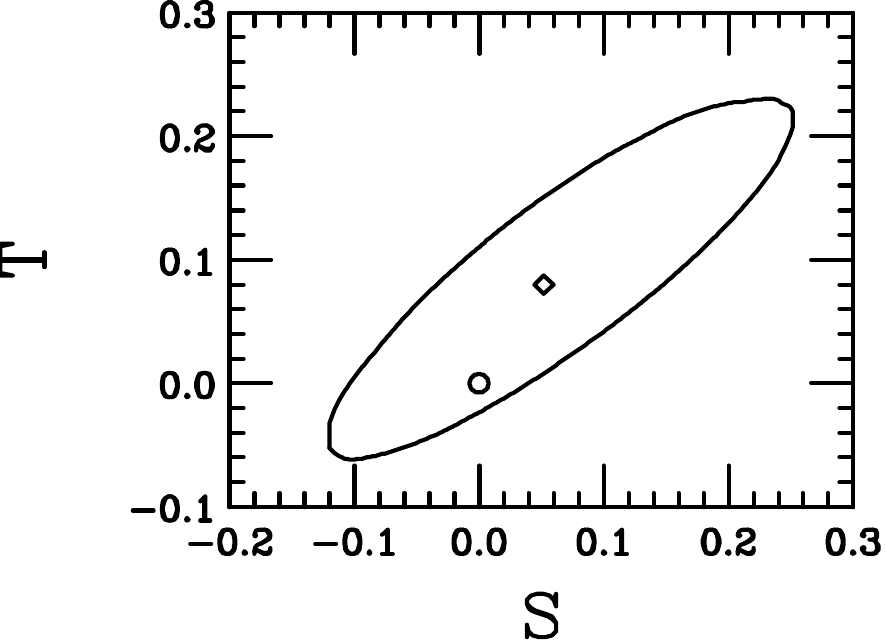}}
\caption{\small{95\% confidence region for $S$ and $T$ with $m_H=125$
    GeV. The diamond indicates the \chisqsp minimum for the
    oblique fit and the circle denotes the Standard Model.}}
\label{fig1}
\end{figure}

Table 1 shows that the tensions are not resolved by oblique new
physics, since the oblique fit has a similar p-value as the SM
fit. Figure 1 displays the 95\% CL contour in the $S-T$ plane, defined
with respect to the best oblique fit, which is at $S,T=
0.05,0.08$. The 95\% limit is at $\chi^2 = 20.3 +5.99= 26.3$. These
results change very little if the low energy data from M\"oller
scattering and atomic parity violation are added to the EWWG data set.
The best SM4 fits approach the \chisqsp and $S,T$ values of the
oblique fit.

\vskip 0.1 in
{\noindent \large \bf{3. SM4 Fits at Two Loop Order}}
\vskip 0.1 in

The SM4 parameter space is five dimensional, with four fermion masses,
$m_T,\, m_B,\, m_N$, $m_L$, and the mixing angle $\theta_{34}$.  We
marginalize over various combinations of the five SM4 parameters and
over the three usual SM3 parameters, $m_t$, $\Delta
\alpha^{(5)}(m_Z)$, $\alpha_S(m_Z)$, with \mhsp fixed at 125
GeV.\footnote{ Because $m_Z$ is known to much greater precision than
  the other SM3 parameters, the \chisqsp and p-values are the same
  whether it is marginalized and constrained or just fixed at its
  experimental central value.} If the SM3 parameters were instead
fixed at their SM3 best fit values, a procedure employed in many fits
of BSM models, we would not obtain the true \chisqsp minimum for the
SM4 model, which typically occurs at different values of the SM3 
parameters than the values in the SM3 fit.

The fits include the nondecoupling, order $(G_Fm_f^2)^2$ two loop
corrections to the $T$ parameter, which are protected by custodial
$SU(2)$ and vanish when the weak doublet partners have equal
mass. They have been calculated in a variety of different
limits; we use the result of van der Bij and
Hoogeveen\cite{vdbh}.  We also use their result for the two loop
hypercharge correction, proportional to $g^{\prime 2}G_Fm_f^2$, which
is important because it does not vanish for equal mass partners, since
hypercharge breaks the custodial $SU(2)$. It is computed for equal
mass partners with $m_Q^2=(m_T^2 + m_B^2)/2$, resulting in a small,
nonleading error, of order $\alpha\, {\rm sin}^2\theta_W/\pi$ times
the one loop term.  Because the hypercharge correction is large and
only begins at two loops, the difference between the one loop
correction and the total two loop correction is not a valid indicator
of the convergence of perturbation theory: the two loop results may
differ substantially from the one loop results even when the
perturbation expansion is under control. We will estimate the
sensitivity of the fits to the uncertainty in the perturbation
expansion by scaling the two loop corrections by the generically
expected uncertainty.

For the nondecoupling contributions to the $S$ parameter, which only
depend logarithmically on the fermion masses, we use the exact one
loop expressions from He, Polonsky, and Su\cite{he_su}. 

In figure 2 we compare one and two loop fits. In the fit on the left
$m_T$ is varied and $m_N$ is marginalized while on the right $m_N$ is
varied and $m_T$ is marginalized. In both cases $m_B=750$ and
$m_L=200$ GeV are fixed with $\theta_{34}=0$.  The neutrino mass is
allowed to vary to the 46 GeV lower limit that applies for stable or
long-lived neutrinos that escape the detector.\footnote{See
  \cite{hm_etal} for a discussion of this scenario.} As a function of
$m_T$ the two loop fit has a lower, broader, and flatter \chisqsp
minimum than the one loop fit and both are approximately symmetric in
$m_T - m_B$. Neutrino masses at the low end of the allowed range are
favored, and the lower \chisqsp of the two loop distribution emerges
primarily at small $m_N$.

\begin{figure}
\centering
\begin{tabular}{cc}
{\includegraphics[width=3in,angle=90]{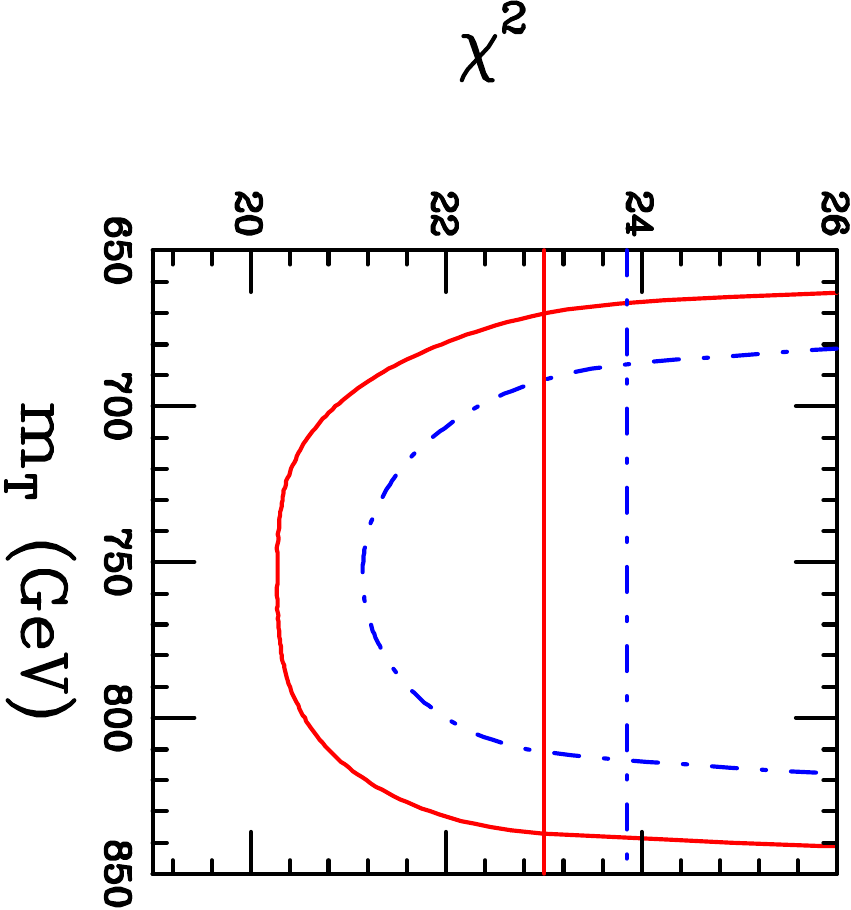}}&
{\includegraphics[width=3in,angle=90]{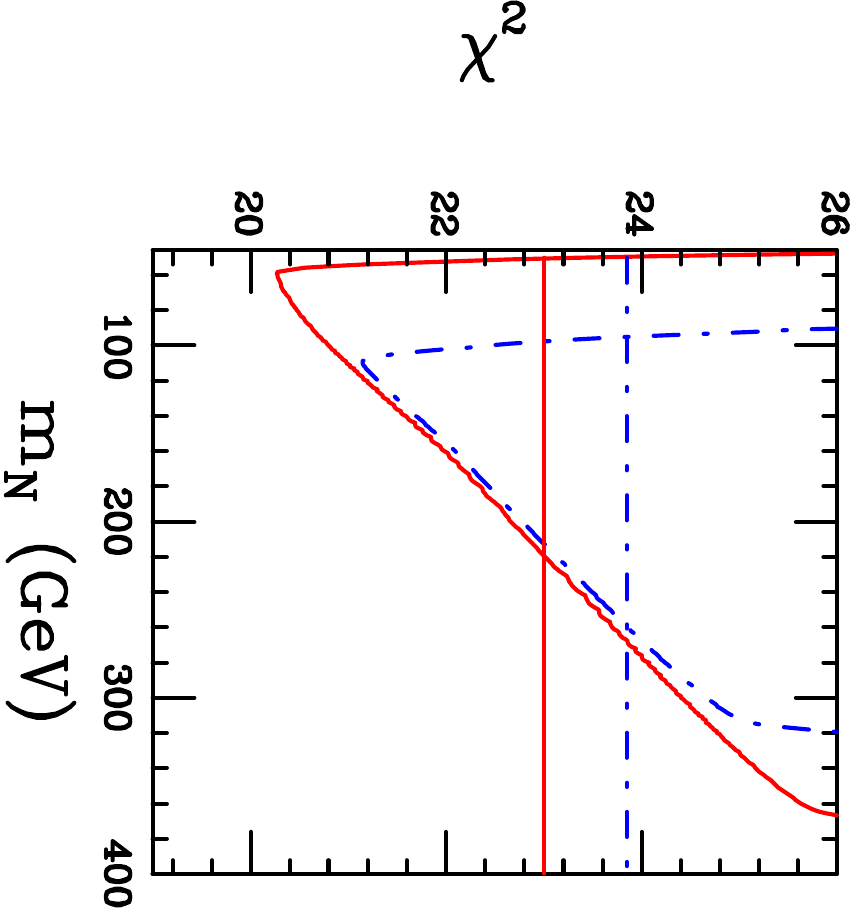}}\\
\end{tabular}
\caption{\small{\chisqsp distributions at one (blue dotdash) and two
    loops (solid red), varying $m_T$ while marginalizing over $m_N$
    (left) and varying $m_N$ while marginalizing over $m_T$
    (right). $m_B=750$ and $m_L=200$ GeV are fixed. The horizontal lines
    indicate the corresponding 90\% confidence intervals.}}
\label{fig2}
\end{figure}

Figure 3 displays the 95\% CL contour plots in the $(m_L - m_N)$ --
$(m_T - m_B)$ plane. In these plots $m_T$ and $m_N$ are marginalized
with $m_L=200$ GeV, $\theta_{34}=0$ and $m_B= 600,750,900$ GeV.  The
one and two loop contours do not overlap, even for $m_B= 600$ GeV, and
become increasingly separated as the quark mass is increased.  The
two loop contours are larger than the one loop contours, as is
apparent in figure 2.

\begin{figure}[b!]
\centering
\begin{tabular}{ccc}
{\includegraphics[width=2in,angle=90]{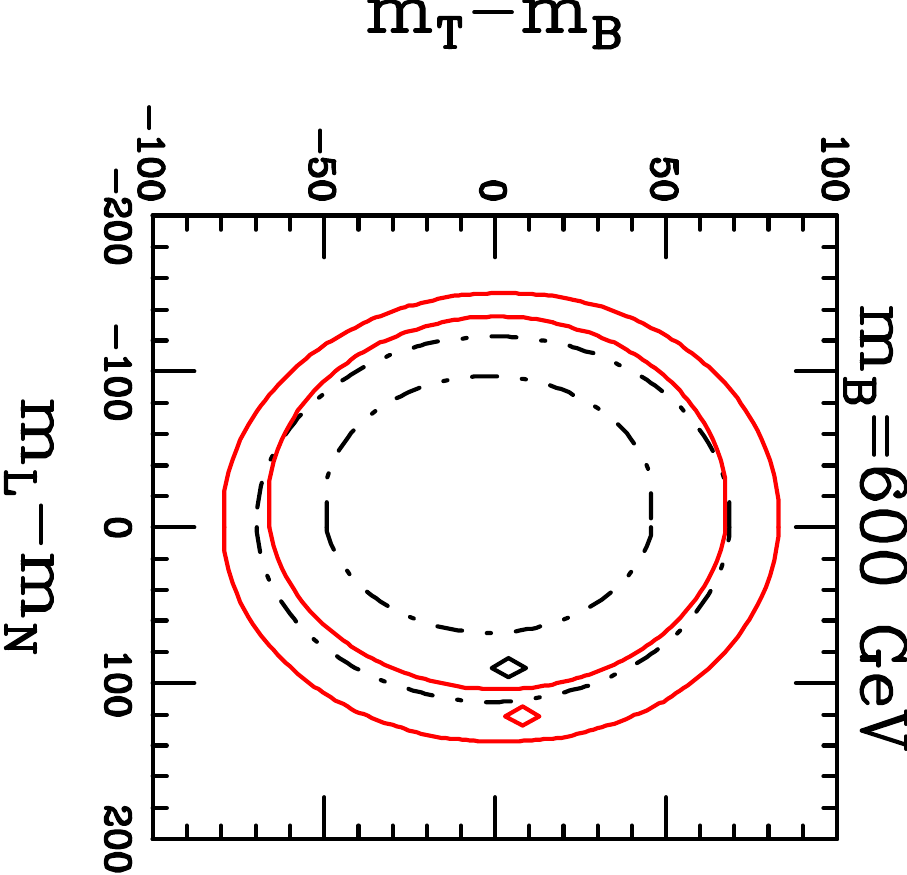}}&
{\includegraphics[width=2in,angle=90]{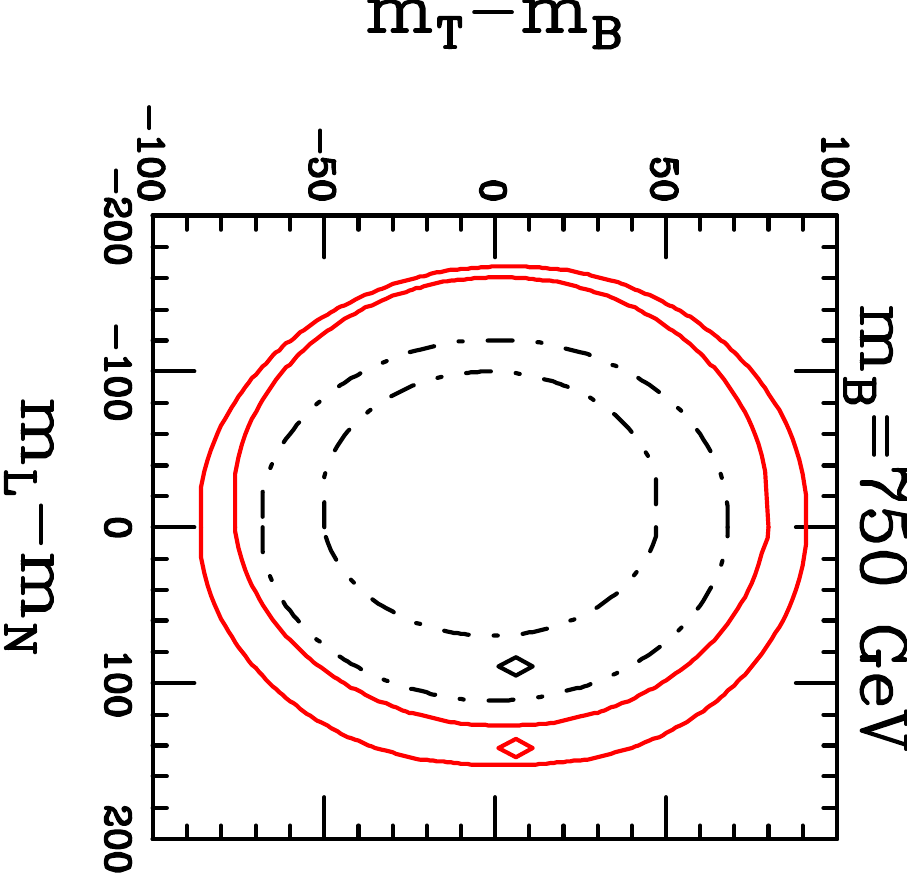}}&
{\includegraphics[width=2in,angle=90]{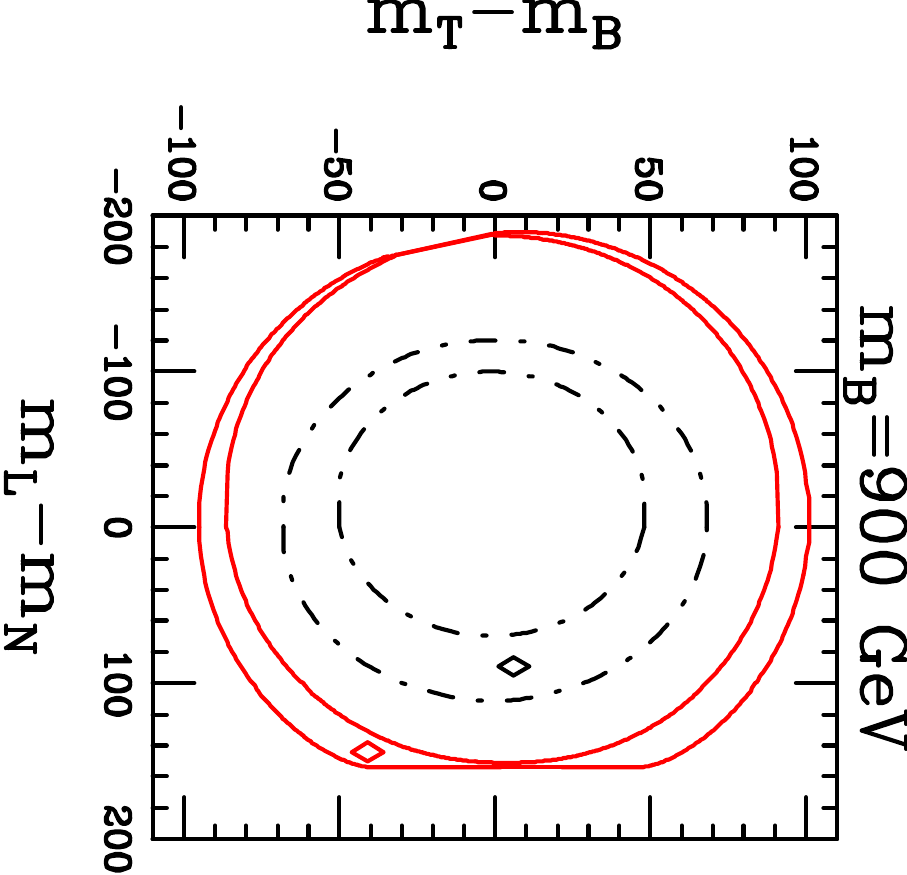}}\\
\end{tabular}
\caption{\small{95\% CL contour plots computed with two loop (solid
    red) and one loop (blue dotdash) corrections with $m_L=200$ GeV,
    $m_B= 600,750,900$ GeV, marginalizing over $m_T$ and $m_N$.
    Diamonds indicate the best fits.}}
\label{fig3}
\end{figure}

The SM4 fits reflect an interplay between the quark and lepton
contributions to $S$ and $T$.  For $\theta_{34}=0$ the leading SM4
corrections are given entirely by $S$ and $T$, so that the oblique fit
shown in table 1 and figure 1 is the limit for how good an SM4 fit can
be. The SM4 fits then choose 4G masses to yield $S$ and $T$ as close
as possible to the the best oblique fit, $S,T = 0.055, 0.08$. 
The quark contribution to $S$, which is rather insensitive to the
specific values of $m_T$ and $m_B$, is large and positive, $S_{TB}
\simeq 0.16$, well above the preferred value. It is offset by the
lepton contribution to $S$ which includes a negative term proportional
to $- {\rm log}(m_L/m_N)$, favoring a large mass difference between
$L$ and $N$ with $m_L > m_N$ as seen in figure 2. However large
$(m_L-m_N)$ induces a large positive contribution to $T$ that can
force $S,T$ outside the 95\% contour in figure 1. The fits then favor
small $T,B$ mass splitting to minimize the quark contribution to $T$,
and the $L,N$ mass difference strikes a balance to achieve the most
negative possible contribution to $S$ while keeping $T$ from becoming
too large. The importance of the two loop quark hypercharge correction
is then apparent: because it makes a large negative contribution to
$T$ it allows $(m_L/m_N)$ to increase further, resulting in fits that more
nearly approach the limiting value of the oblique fit.

\begin{table}[t!]
\begin{center}
\vskip 12pt
\begin{tabular}{c|c|c|c||c}
\hline
\hline
{\boldmath $m_B$} & {\bf 600} & {\bf 750}& {\bf 900}& {\boldmath $750:\ m_T$ {\bf [95\%]}} \\
\hline
{\boldmath $m_T$} & 608 & 756 & 859 & 841.4\\
{\boldmath $m_N$} & 79  & 59  & 56 & 205.5 \\
\hline
$S_{LN}$ & -0.055 & -0.10 & -0.109 & +0.057 \\
$T_{L-N}^{(1)}$ & 0.25 & 0.334 & 0.35 & 0.00054\\
$T_{L-N}^{(2)}$ & 0.0019 & 0.0018 & 0.0017 & 0.0000071\\
$T_{LN}^{({\rm Y})}$ & -0.00026 & -0.00019 & -0.00018 & -0.0015\\
$T_{LN}^{({\rm TOT})}$ & 0.25 & 0.336 & 0.35 & -0.00094 \\
\hline
$S_{TB}$ & 0.158 & 0.16 & 0.164 & 0.147\\
$T_{T-B}^{(1)}$ & 0.0034 & 0.0019 & 0.090 & 0.45 \\
$T_{T-B}^{(2)}$ & 0.00040 & 0.00035 & 0.023 & 0.099\\
$T_{TB}^{({\rm Y})}$ & -0.14 & -0.256 & -0.38 & -0.30\\
$T_{TB}^{({\rm TOT})}$ & -0.14 & -0.254 & -0.27 & +0.25\\
\hline
$S_{\rm TOT}$ & 0.10 & 0.06 & 0.055 & 0.20\\
$T_{\rm TOT}$ & 0.11 & 0.08 & 0.08 & 0.25\\
\hline
\hline
\end{tabular}
\end{center}
\caption{\small{First three columns: contributions to $S$ and $T$ for
    the best fits with $m_L=200$ and $m_B= 600,750, 900$ GeV,
    marginalizing over $m_N$ and $m_T$. Fourth column: $m_L=200$ and
    $m_B= 750$ GeV with $m_T$ at its 95\% CL upper limit. }}
\end{table}

These features are visible in table 2, which dissects the quark and
lepton contributions to $S$ and $T$. For each fit $m_L=200$ GeV is
fixed. The first three columns are for the best fits with $m_B=
600,750, 900$ GeV and the fourth is for $m_B= 750$ GeV with $m_T$ at
its 95\% upper limit. For the three best fits the two loop hypercharge 
correction makes the dominant contribution to $T$, especially for 
$m_B= 600$ and 750 GeV where the one loop quark contribution 
is negligible in comparison. At the 95\% upper limit on $m_T$, in 
the fourth column, the hypercharge correction is comparable to 
although smaller than the one loop term. The best fit at $m_B= 900$ 
GeV reaches the $S,T$ values of the oblique fit. 

\vskip 0.1 in
{\noindent \large \bf{4. Convergence of the Perturbation Expansion}}
\vskip 0.1 in

In this section we examine the sensitivity of the fits to uncertainty
in the perturbation expansion, which increases with increasing quark
mass. Because of the large contribution from the hypercharge
correction, which only begins at two loops, the validity of the
expansion cannot be judged simply by the difference between the one
and two loop fits.

Consider for instance the fit with $m_B=600$ GeV in table 2.  The one
loop result, $T^{(1)}=0.25$, differs by more than a factor two from
the total result at two loops , $T_{\rm TOT}=0.11$, but this difference
tells us nothing about the convergence of the loop expansion.  The one
loop result is completely dominated by the leptonic contribution, for
which perturbation theory is certainly reliable since $m_L$ and $m_N$
are well within the perturbative domain, as is explicitly evident from
the negligible value of the leptonic two loop contributions to $T$ in
the table. The quark contribution to $T$ is completely dominated by
the quark hypercharge term, which is large because it is not
suppressed by the near degeneracy of the $T$ and $B$ masses favored by
the fits.  

To assess the reliability of the expansion we need the next order
contribution to the quark hypercharge term, a three loop correction
that is not known.  The best we can do is to use a conservative
generic estimate of the loop expansion parameter, $~ y_Q^2/4\pi^2$,
where $y_Q=m_Q/v$ is the quark coupling to the Higgs boson and $v=247$
GeV. In particular we use the known ratio of the one and two loop
quark corrections\cite{vdbh} that result from the mass difference of
$T$ and $B$ for the case $|m_T - m_B| \ll m_B$, which is consistent
with and just a factor 3/4 smaller than the generic estimate,
$$ 
R_{12}={{T_{T-B}^{(2)}}\over {T_{T-B}^{(1)}}} = {3\over 16
  \pi^2}{m_Q^2\over v^2}.
$$ 
We then have $R_{12}= 0.11, 0.18, 0.25$ for $m_Q= 600, 750, 900$ GeV.

To exhibit the effect of uncertainty of this magnitude we
compare fits that shift the hypercharge correction by 
a factor $\pm R_{12}$,
$$
T_{TB}^{({\rm Y})} \to T_{TB}^{({\rm Y})}\cdot (1 \pm R_{12}).
$$ 
Figure 4 compares \chisqsp distributions with
$T_{TB}^{({\rm Y})}$ rescaled as above for
$m_B=750$ GeV and $m_L=200$ GeV, as in figure 2. The distributions 
are very similar over most of the allowed range in $m_T$ and $m_N$,
with significant differences only near the upper and lower limits on $m_T$
and the lower limit on $m_N$.  The rescaling of the hypercharge
correction shifts the 95\% limits on $m_T$ and $m_N$ by $~5$ GeV or
less.

\begin{figure}
\centering
\begin{tabular}{cc}
{\includegraphics[width=3in,angle=90]{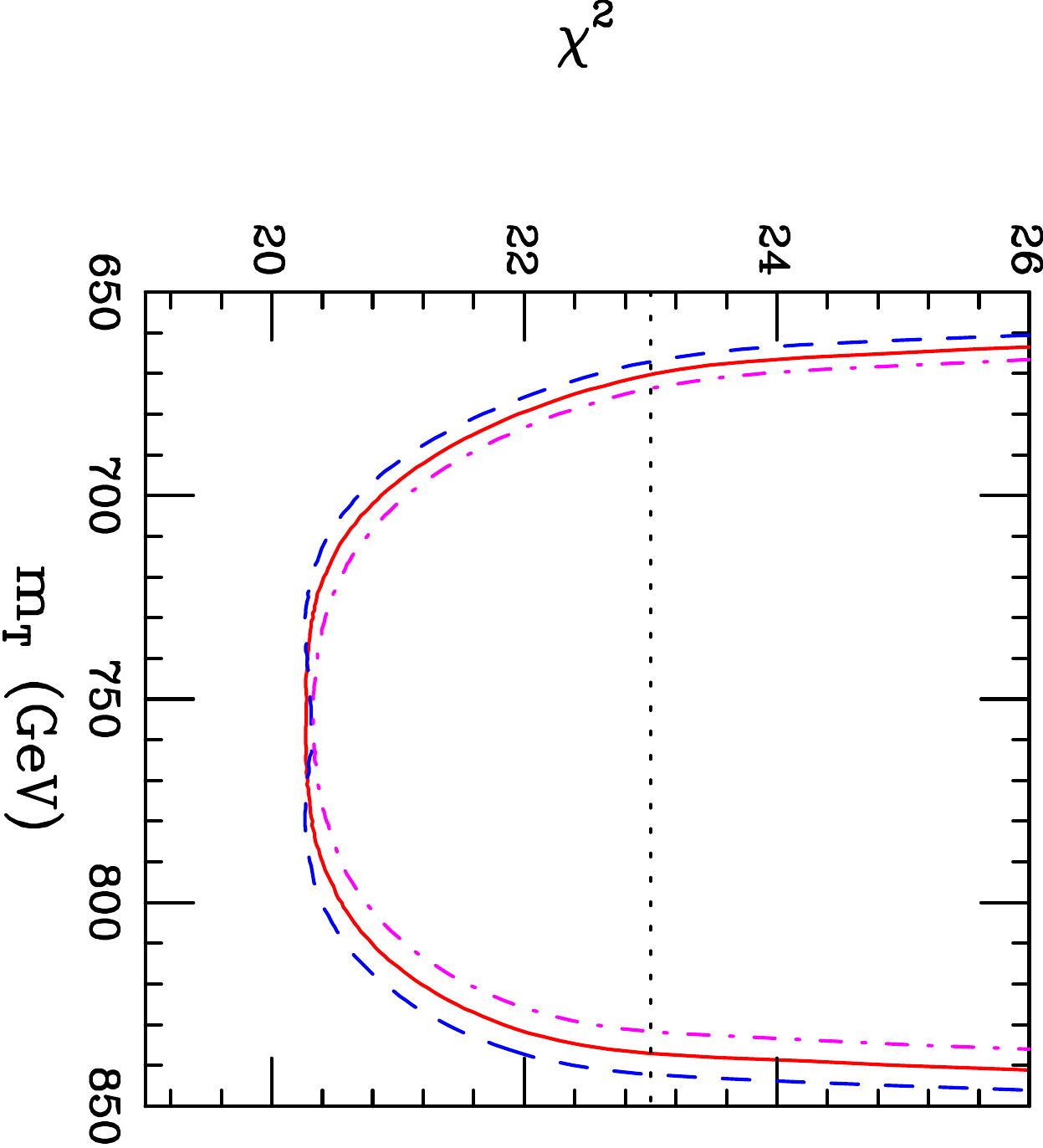}}&
{\includegraphics[width=3in,angle=90]{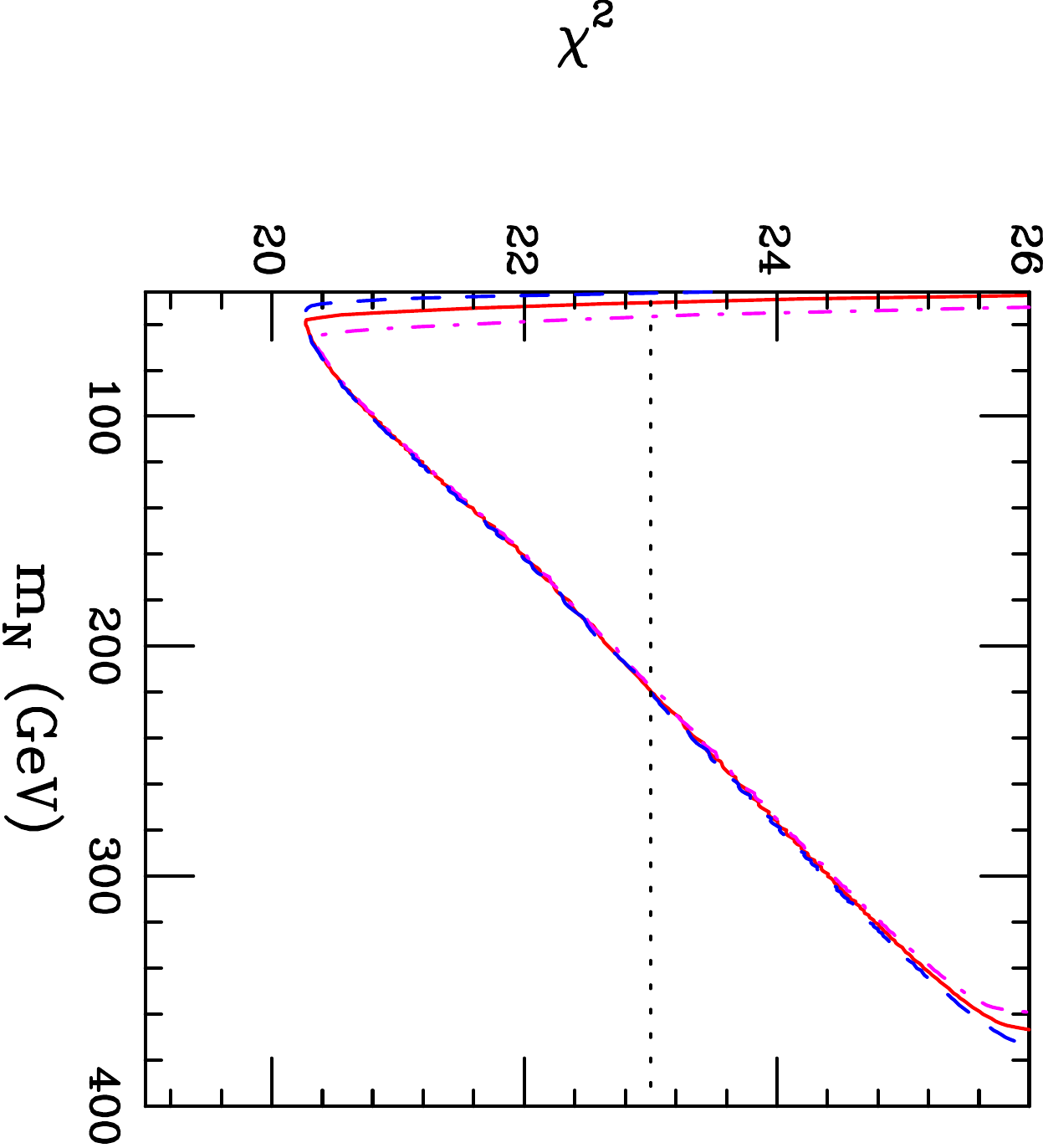}}\\
\end{tabular}
\caption{\small{\chisqsp distributions computed at two loops with
    $T_{TB}^{({\rm Y})}$ rescaled by $1+R_{12}$ (blue dashes) and
    $1-R_{12}$ (magenta dotdash), compared with the central value
    (solid red). $m_B=750$ and $m_L=200$ GeV are fixed. The dotted
    lines mark the 90\% confidence intervals.}}
\label{fig4}
\end{figure}

\begin{figure}[b!]
\centering
\begin{tabular}{ccc}
{\includegraphics[width=2in,angle=90]{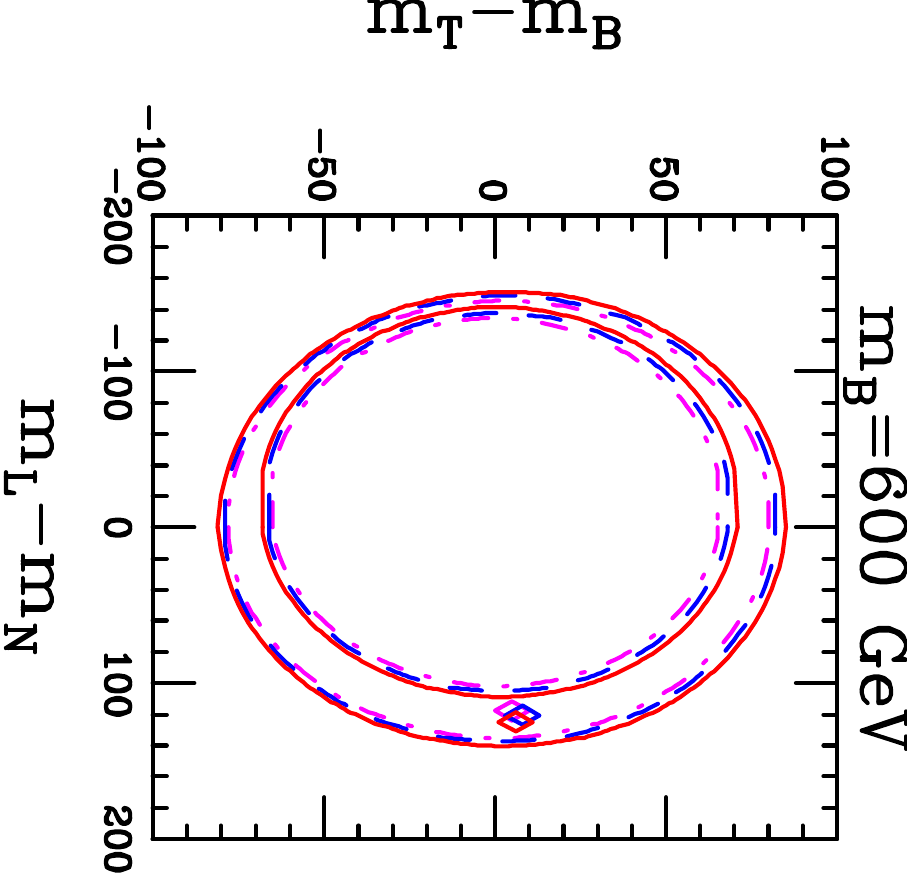}}&
{\includegraphics[width=2in,angle=90]{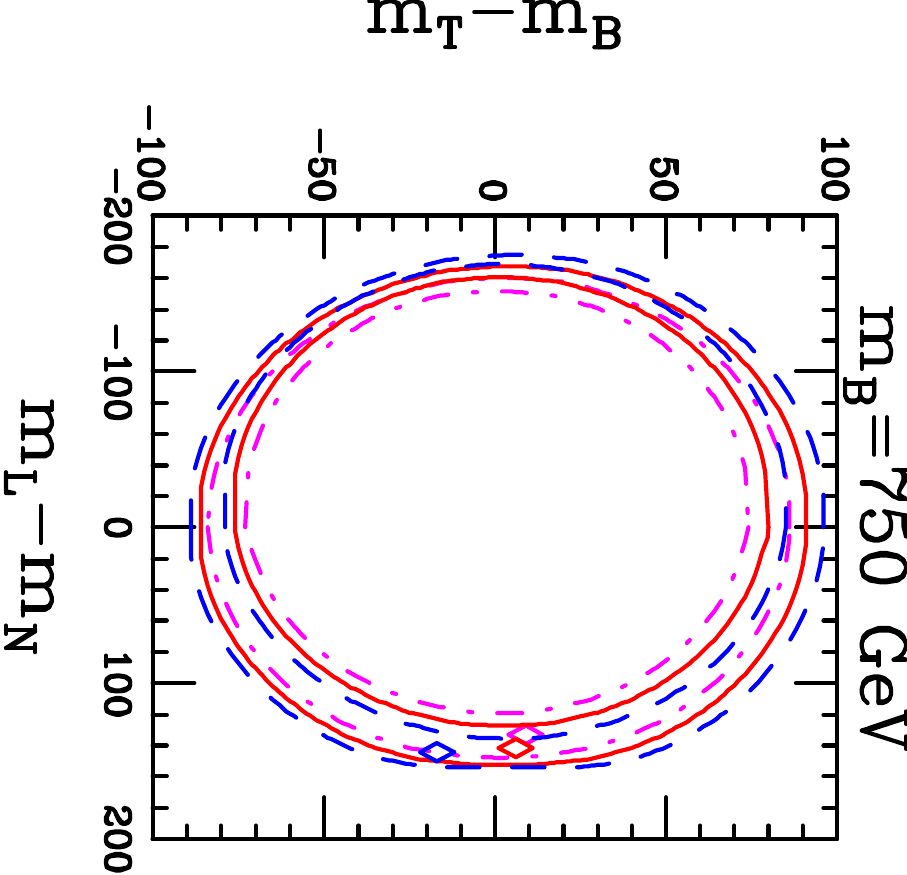}}&
{\includegraphics[width=2in,angle=90]{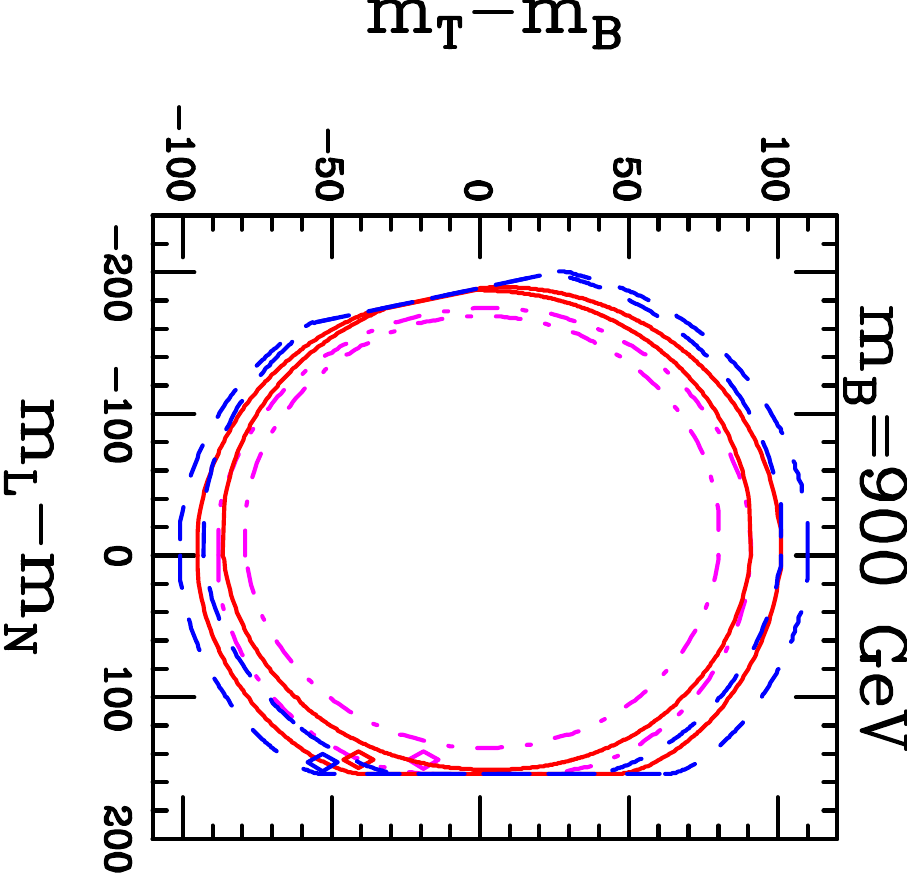}}\\
\end{tabular}
\caption{\small{95\% CL contour plots , marginalizing over $m_T$ and
    $m_N$, computed at two loops with $T_{TB}^{({\rm Y})}$ rescaled by
    $1+R_{12}$ (blue dashes) and $1-R_{12}$ (magenta dotdash),
    compared with the central value (solid red). The charged lepton
    mass is fixed at $m_L=200$ GeV.  Diamonds indicate the best
    fits.}}
\label{fig5}
\end{figure}

\begin{figure}[t!]
\centering
\begin{tabular}{ccc}
{\includegraphics[width=2in,angle=90]{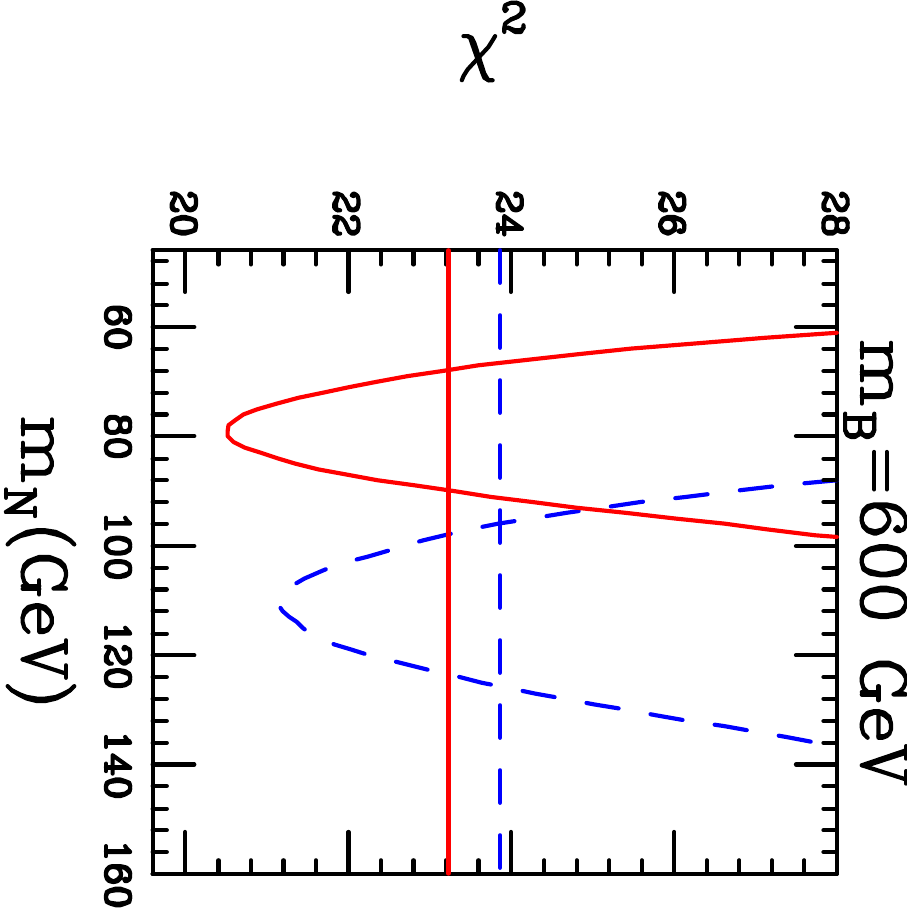}}&
{\includegraphics[width=2in,angle=90]{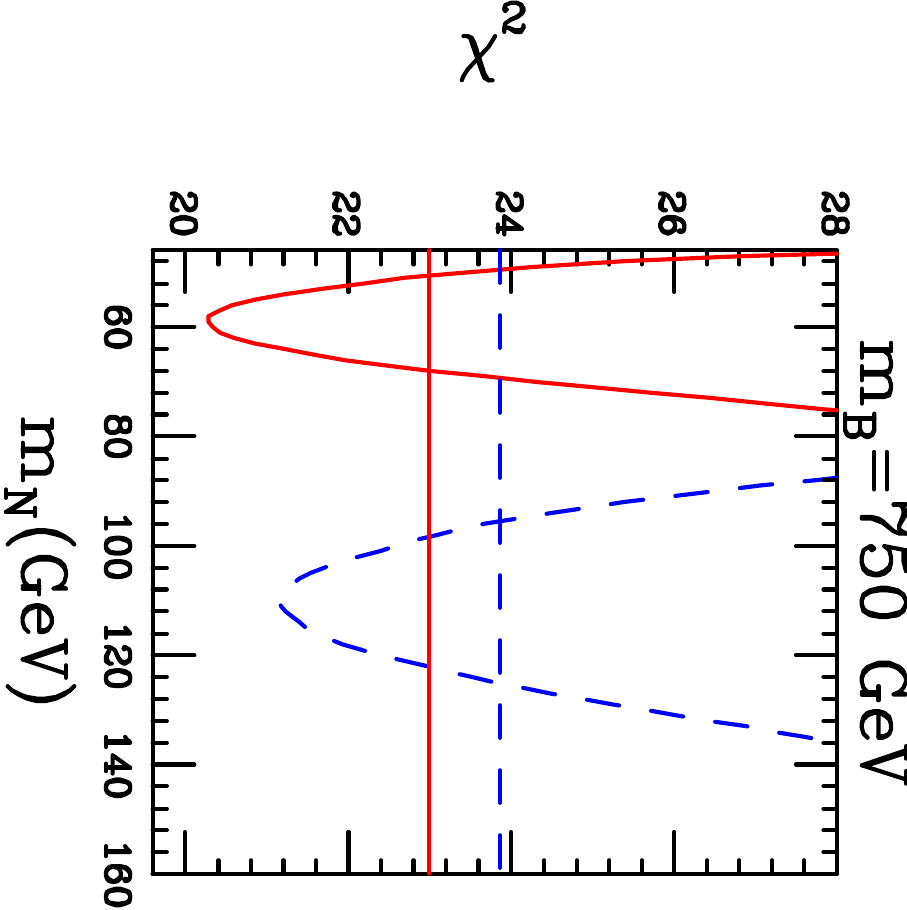}}&
{\includegraphics[width=2in,angle=90]{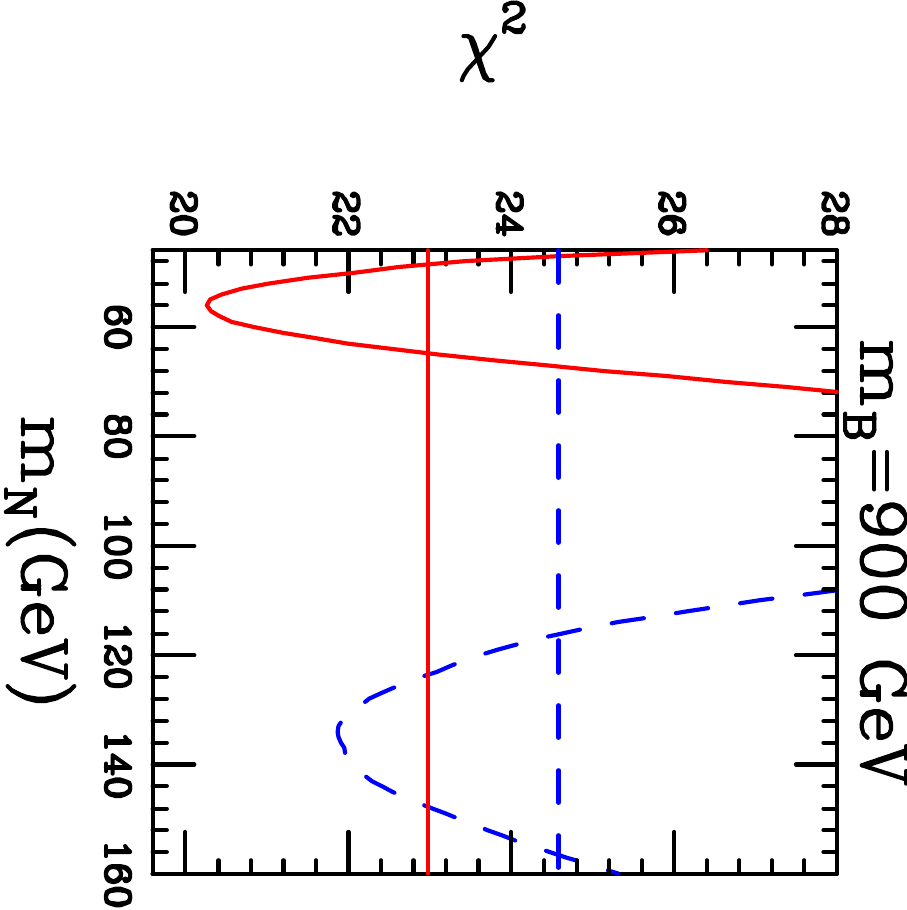}}\\
\end{tabular}
\caption{\small{\chisqsp distributions as a function of $m_N$ with
    $m_L=200$ GeV and $m_T$ chosen as described in the text. The
    dashed blue and red distributions are from one and two loop fits
    respectively. The intersections of the distributions with the
    corresponding horizontal lines define the 90\% confidence
    intervals for $m_N$.}}
\label{fig6}
\end{figure}

Similarly, the contour plots in the $(m_L-m_N)$ -- $(m_T-m_B)$
plane are compared in figure 5, for $m_L=200$ GeV and
$m_B=600,750,900$ GeV, as in figure 3. For $m_B=600$ GeV the three
contours overlap quite closely, even though 600 GeV exceeds the tree
unitarity perturbative limit. For $m_B=750$ GeV they begin to diverge
noticeably, as expected from the \chisqsp distributions in figure
4. Finally, for $m_B=900$ GeV the three 95\% contours are almost
completely non-overlapping, suggesting that perturbation theory may be
of limited value at this mass scale.

\begin{figure}[b!]
\centering
\begin{tabular}{ccc}
{\includegraphics[width=2in,angle=90]{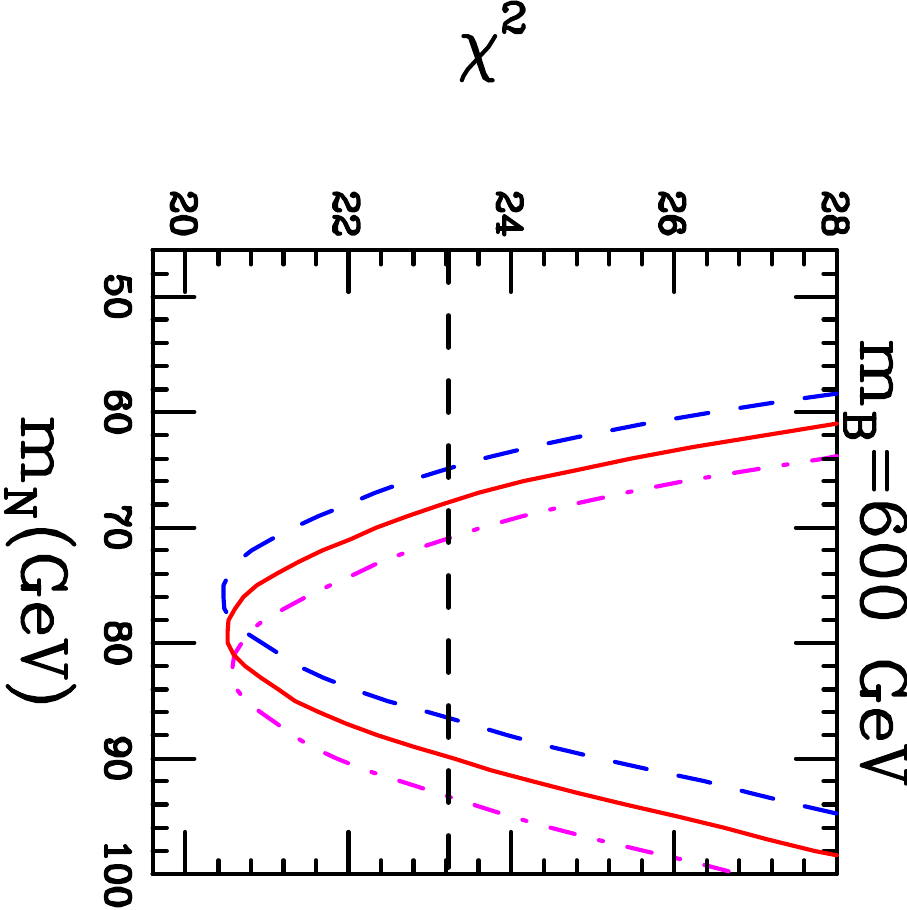}}&
{\includegraphics[width=2in,angle=90]{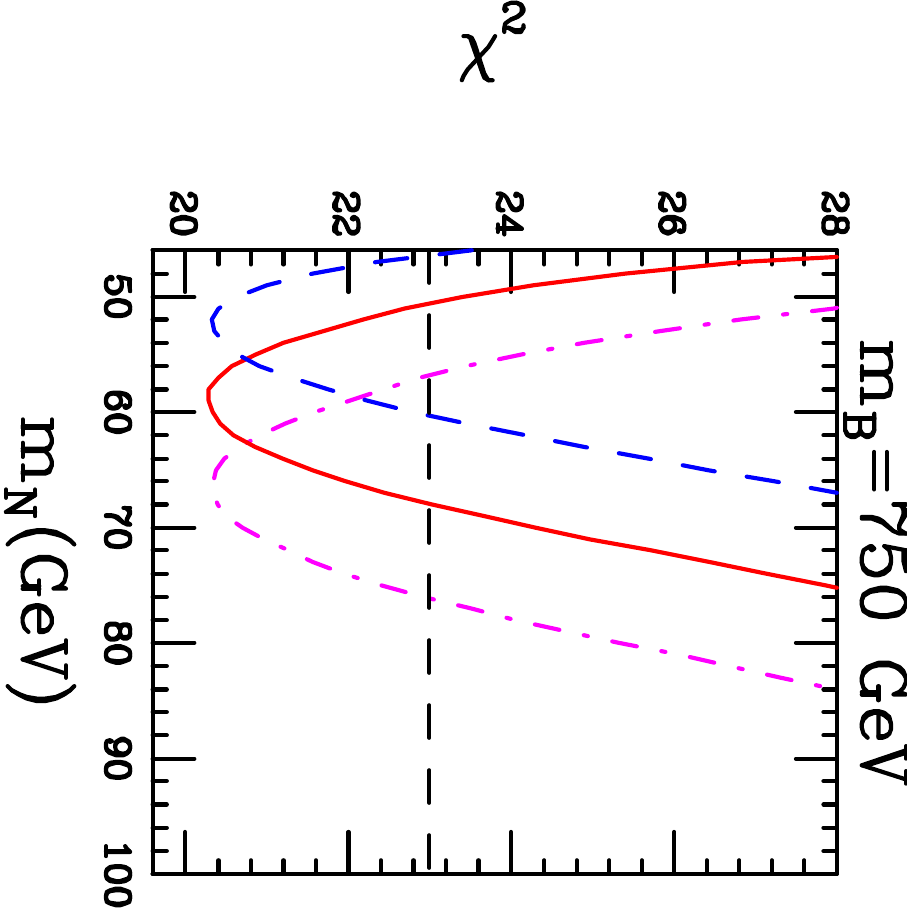}}&
{\includegraphics[width=2in,angle=90]{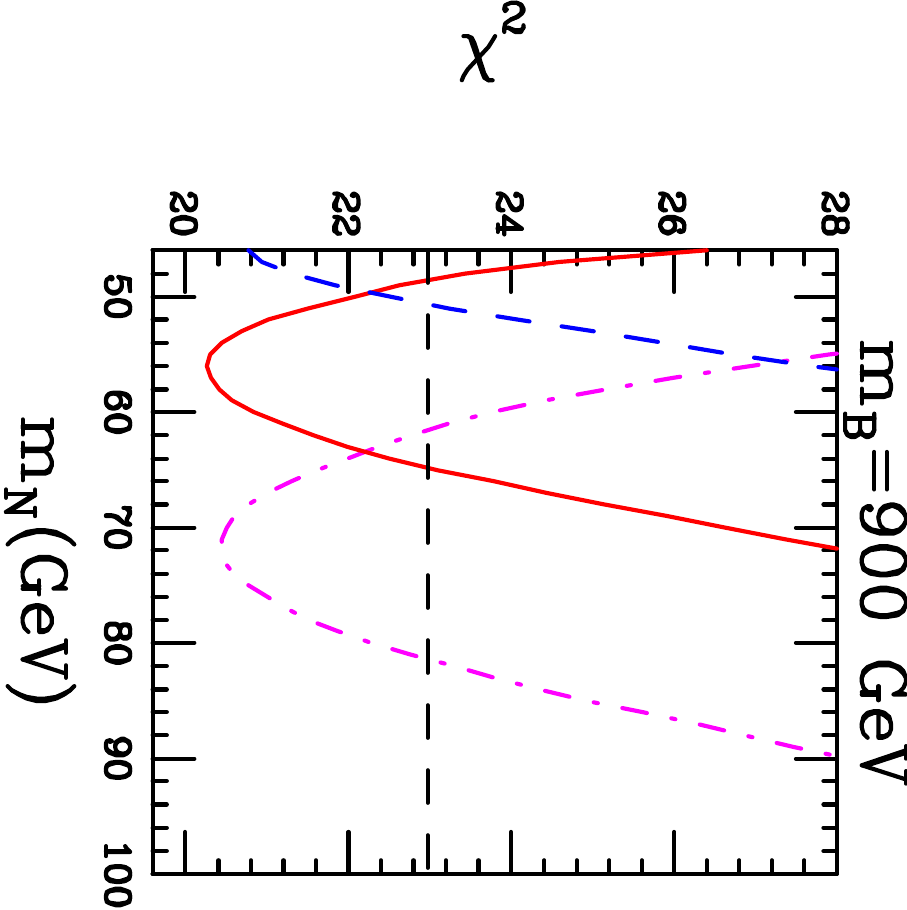}}\\
\end{tabular}
\caption{\small{Two loop \chisqsp distributions as a function of $m_N$
    with $m_L=200$ GeV and $m_T$ chosen as described in the text.  The
    hypercharge correction $T_{TB}^{({\rm Y})}$ is rescaled by
    $1+R_{12}$ (blue dashes) and $1-R_{12}$ (magenta dotdash),
    compared with the central value (solid red). The horizontal lines
    indicate the 90\% confidence intervals.}}
\label{fig7}
\end{figure}

Finally we consider an example of how the electroweak corrections will
be used in practice if evidence of a fourth generation is discovered
at the LHC. We imagine that the $T$ and $B$ quarks and the charged
lepton $L$ are discovered and their masses measured, and we consider
how well the electroweak fit can then constrain the mass of the yet
undiscovered heavy neutrino $N$. In figure 6 we compare the
constraints for one and two loop fits, and in figure 7 we compare the
two loop fits with $T_{TB}^{({\rm Y})}$ smeared as described above. We
assume $m_L=200$ GeV and $m_B=600,750,900$ GeV. In each case $m_T$ is
fixed at its value for the \chisqsp minimum obtained by marginalizing
over $m_T$ and $m_N$, and \chisqsp is then obtained as a function of 
$m_N$. 

\begin{table}
\begin{center}
\vskip 12pt
\begin{tabular}{c|c|c}
\hline
\hline
{\boldmath $m_B$} & {\bf Unsmeared} & {\bf Smeared} \\
\hline
600 & 68,90 & 65,93 \\
750 & 51,68 & 46,76 \\
900 & 48,65 & 46,81 \\
\hline
\hline
\end{tabular}
\end{center}
\caption{\small{ 90\% confidence intervals for $m_N$ in GeV, as in 
figure 7. The unsmeared values are from the central (red) distributions 
in figure 7, while the smeared values are the outer envelopes of 
the distributions obtained by rescaling the hypercharge correction 
as described in the text.}}
\end{table}

We see in figure 6 that the 90\% confidence intervals for $m_N$ of the
one and two loop fits are completely disjoint, even for $m_B=600$ GeV,
with increasing separation for larger $m_B$. In figure 7 we see that
smearing the hypercharge correction by $1 \pm R_{12}$ results in
substantially overlapping confidence intervals for $m_B=600$ GeV that
become almost completely disjoint for $m_B=900$ GeV. Table 3 shows the
effect of the smearing on the 90\% confidence intervals. For
$m_B=600$ the impact of the smearing is modest, adding 3 GeV to the
upper and lower limits. For $m_B=900$ GeV the effect is appreciable,
with the 95\% CL upper limit on $m_N$ increasing from 65 to 81 GeV.  A
comparable change would occur for the lower limit but is
precluded by the direct lower limit on $m_N$ at 46 GeV.

\vskip 0.1 in
{\noindent \large \bf{5. Discussion}}
\vskip 0.1 in

If the Higgs-like particle at 125 GeV is confirmed as {\em the} Higgs
boson of the standard model, then SM4 appears to be excluded, although
BSM4 variants with additional new quanta might still be viable. If it
is a non-SM Higgs boson, e.g., a denizen of a 2HDM, then 4G models can
also still be viable. We have used SM4 as a laboratory to study the
effect of two loop corrections on the EW fit for 4G models. The
results are qualitatively applicable to the broader class of 4G
models, but detailed results can only be obtained from the examination
of each particular model. For instance, for 2HDM models in regions of
the parameter space with enhanced Yukawa coupling, the effect of the
two loop hypercharge corrections could be even larger than the already
large effect we have found in SM4.

In the SM4 ``laboratory'' we have identified an important two loop
correction that has a big effect on the EW fit in 4G models. Although
computed 25 years ago by van der Bij and Hoogeveen, it has not
previously been included in EW fits of 4G models. It is important
because it is the first correction that breaks the custodial $SU(2)$
even if weak doublet partners have equal mass, and, because it makes a
large negative contribution to the rho parameter, it allows the fits
to approach more closely to the limit of the oblique fit in which
$S,T$ are free parameters. It is then important in the region of
parameter space preferred by the EW fits, where the $T$ and $B$ quarks
have nearly equal masses so that the one loop quark correction is
small. In that experimentally preferred region it is by far the
largest correction to the rho parameter from the 4G quark
sector. Because of the hypercharge correction the two loop fits differ
significantly from the one loop fits even when the quark masses are
light enough that perturbation theory is reasonably convergent.

We have also studied the convergence of the perturbation expansion
with increasing quark mass and its effect on the 4G contraints from
the EW data, using a generic estimate of the order of magnitude of the
three loop hypercharge correction, which is the largest unknown term
in the parameter region preferred by the fits. We find that for
$m_Q=600$ GeV perturbation theory provides useful guidance, even
though tree unitarity for elastic $\overline QQ$ scattering is
saturated at at $m_Q=500$ GeV. For $m_Q= 900$ GeV the estimated
uncertainty increases so that it undermines the usefulness of the
expansion. To obtain better etimates of the convergence of
perturbation theory, it would be necessary to compute the the three
loop hypercharge correction in the relevant models. Motivation to meet
this challenge could be found if (and probably only if) evidence
emerges for the existence of a fourth generation.

\vskip .2in
\noindent{\small This work was supported in part by the Director, Office of
Science, Office of High Energy and Nuclear Physics, Division of High
Energy Physics, of the U.S. Department of Energy under Contract
DE-AC02-05CH11231}


\vskip .2in

\begin{thebibliography}{99}

\bibitem{hashimoto}M.~Hashimoto, Phys.Rev.{\bf D81}, 075023 (2010).
arXiv:1001.4335

\bibitem{holdom_lhp}B.~Holdom, Nuovo Cim. {\bf C035N3}, 71 (2012).
arXiv:1201.0074

\bibitem{lenz_no} O.~Eberhardt , G.~Herbert, H.~Lacker, A.~Lenz,
  A.~Menzel {\it et al.}, Phys.Rev.Lett. {\bf 109}, 241802 (2012),
  arXiv:1209.1101 (2012). For analyses using earlier data sets see
  also O.~Eberhardt, A.~Lenz, A.~Menzel, U.~Nierste, and M.~Wiebusch,
  Phys.Rev.{\bf D86}, 074014 (2012), arXiv:1207.0438; O.~Eberhardt,
  G.~Herbert, H.~Lacker, A.~Lenz, A.~Menzel, {\it et al.},
  Phys.Rev.{\bf D86}, 013011 (2012); A.~Djouadi and A.~Lenz,
  Phys.Lett.{\bf B715}, 310 (2012), arXiv:1204.1252; E.~Kuflik,
  Y.~Nir, T.~Volansky(2012), arXiv:1204.1975. Crucial NLO corrections
  were obtained in G.~Passarino, C.~Sturm, S.~Uccirati, Phys.Lett.{\bf
    B706} 195 (2011), arXiv:1108.2025; and in A.~Denner {\it et al.},
  Eur.Phys.J.{\bf C72} 1992 (2012), arXiv:1111.6395.

\bibitem{4G_virtues} B.~Holdom, W.S.~Hou {\it et al.}, PMC Phys.{\bf
  A3}, 4 (2009), arXiv:0904.4698; P.~Frampton, P.Q.~Hung, M.~Sher,
  Phys.Rept.{\bf 330}, 263 (2000).

\bibitem{soni}M.~Geller, S.~Bar-Shalom, G.~Eilam, and A.~Soni,
  arXiv:1209.4081 (2012); S.~Bar-Shalom, M.~Geller, S.~Nandi, and
  A.~Soni, arXiv:1208.3195 (2012); S.~Bar-Shalom, S.~Nandi, and
  A.~Soni, Phys.Rev.{\{\bf D84}, 053009 (2011).

\bibitem{he-chen} N.~Chen and H.-J.~He, JHEP {\bf 1204}, 062 (2012),
arXiv:1202.3072.

\bibitem{he-valencia} X.-G.~He and G.~Valencia, Phys.Lett. {\bf B707},
  381 (2012), arXiv:1108.0222.

\bibitem{2dhmG4_EW} L.~Bellantoni {\it et al.}, Phys.Rev. {\bf D86}
  034022 (2012).

\bibitem{mchm} See for instance K.~Agashe, R.~Contino, A.~Pomarol,
  Nucl.Phys. {\bf B719} 165, (2005), arXiv:hep-ph/0412089; G.~Giudice,
  C.~Grojean, A.~Pomarol, R.~Rattazzi, JHEP {\bf 0706} 045 (2007),
  arXiv:hep-ph/0703164; B.~Bellazzini {\it et al.}, arXiv:1205.4032
  [hep-ph] (2102).

\bibitem{pq123} C.M.~Ho, P.Q.~Hung, T.W.~Kephart, JHEP {\bf 1206}, 045
  (2012), arXiv:1102.3997; P.Q.~Hung, C.~Xiong, Phys.Lett. {\bf B694},
  430 (2011), arXiv:0911.3892; P.Q.~Hung, C.~Xiong, Nucl.Phys. {\bf
    B847}, 160 (2011), arXiv:0911.3890.

\bibitem{pq4} P.Q.~Hung, C.~Xiong, Nucl.Phys. {\bf B848}, 288 (2011),
  arXiv:1012.4479.

\bibitem{dilaton_refs} W.D.~Goldberger, B.~Grinstein, Phys.Rev.Lett.
  {\bf 100}, 111802 (2008); Z.~Chacko, R.K.~Mishra, arXiv:1209.3022
  (2012); Z.~Chacko, R.~Franceschini, R.K.~Mishra, arXiv:1209.3259
  (2012); B.~Bellazzini {\it et al.}, arXiv:1209.3299 (2012); T.~Abe
  {\it et al.}, arXiv:1209.4544 (2012); S.~Matsuzaki, K.~Yamawaki,
  Phys.Rev. {\bf D85} 095020 (2012), arXiv:1201.4722.

\bibitem{ew_sm4} H.J.~He, N.~Polonsky, and S.F.~Su,
  Phys.Rev. {\bf D64}, 053004 (2001); V.A.~Novikov, L.B.~Okun,
  A.N.~Rozanov, and M.I.~ Vysotsky, Phys. Lett. {\bf B529}, 111
  (2002); Pis’ma Zh.  Eksp. Teor. Fiz. {\bf 76}, 158 (2002); JETP
  Lett. {\bf 76}, 127 (2002); G.D.~Kribs, T.~Plehn, M.~Spannowsky, and
  T.M.P.~Tait, Phys. Rev. {\bf D76}, 075016 (2007).

\bibitem{ew_ckm4} M.S.~Chanowitz, Phys. Rev. {\bf D79}, 113008 (2009).


\bibitem{radion} M.~Frank, B.~Korutlu, M.~Toharia, Phys.Rev. {\bf D85}
  115025 (2012), arXiv:1204.5944.

\bibitem{cmsB} CMS Collaboration, CMS-PAS-EXO-11-036,
  arXiv:submit/0449821 [hep-ex] (2011).

\bibitem{atlasT} ATLAS Collaboration (Georges Aad {\it et al.}),
  arXiv:1210.5468 (2012).

\bibitem{stop_limit} CMS Collaboration (S.~Chatrchyan {\it et al.}), 
Phys.Lett. {\bf B713}, 408 (2012), arXiv:1205.0272.

\bibitem{madgraph} J.~Alwall {\it et al.}, JHEP {\bf 1106}, 128
  (2011), arXiv:1106.0522.

\bibitem{QQ_Kfactor} J.M.~Campbell, R.~Fredrix, F.~Maltoni,
  F.~Tramontano, JHEP {\bf 0910}, 042 (2009), arXiv:0907.3933;
  E.L.~Berger, Q-H~Cao, Phys. Rev. {\bf D81}, 035006 (2010),
  arXiv:0909.3555.

\bibitem{bound_states} K.~Ishiwata, M.B.~Wise, Phys.Rev. {\bf D83}
  074015 (2011) arXiv:1103.0611; T.~Enkhbat, W-S~Hou, H.~Yokoya,
  Phys.Rev. {\bf D84}, 094013 (2011), arXiv:1109.3382.

\bibitem{alt_decays} P.Q.~Hung, M.~Sher, Phys. Rev. {\bf D77}, 037302
  (2008); C.J.~Flacco, D.~Whiteson, T.M.P.~Tait, S.~Bar-Shalom
  (Technion), Phys.Rev.Lett. {\bf 105},111801 (2010), arXiv:1005.1077.

\bibitem{cfh}  M.S.~Chanowitz, M.A.~Furman, I.~Hinchliffe,
Phys.Lett.{\bf B78}:285,1978; Nucl.Phys.{\bf B153}:402,1979.

\bibitem{veltman} M.~Veltman, Nucl. Phys. {\bf B123}, 89 (1977).

\bibitem{pt} M.~Peskin and T.~Takeuchi, Phys. Rev. {\bf D46}, 381 (1992).

\bibitem{vdbh} J.~Van der Bij and F.~Hoogeveen, Nucl. Phys. {\bf
  B283}, 477 (1987).

\bibitem{new_rb} A.~Freitas, Y.-C.~Huang, JHEP {\bf 1208} 050 (2012),
  arXiv:1205.0299.

\bibitem{ewwg} The ALEPH, DELPHI, L3, OPAL, SLD Collaborations, the
  LEP Electroweak Working Group, the SLD Electroweak, and Heavy
  Flavour Groups, Phys. Rep. {\bf 427}, 257 (2006). See
  http://lepewwg.web.cern.ch/ LEPEWWG/ for the most recent data.

\bibitem{zfitter} A.~B. Arbuzov {\it et al}.,
  Comput. Phys. Commun. {\bf 174}, 728 (2006).

\bibitem{msc_ggi} M.~Chanowitz, talk posted at
  http://www.ggi.fi.infn.it/talks/talk1452.pdf (2010).

\bibitem{ltw-rb} B.~Batell, S~Gori, L.-T.~Wang, arXiv:1209.6382.

\bibitem{he_su} H.-J. He, N. Polonsky, and S.-F. Su, Phys. Rev. {\bf D64},
  053004 (2001).

\bibitem{hm_etal} H.~Murayama {\it et al.}, Phys.Lett. {\bf B705} 208
  (2011),  arXiv:1012.0338.


\end{thebibliography}
\end{document}